\documentclass[pre,amsfonts,amssymb,eqsecnum,floatfix]{revtex4}
\pdfoutput=1
\usepackage{amsmath}
\usepackage{graphicx}
\usepackage{bm}
\usepackage{color}
\allowdisplaybreaks

\begin{document}	
\title{Stochastic processes subject to a reset-and-residence mechanism: transport properties and first arrival statistics}
\author{Axel Mas\'o-Puigdellosas}
\author{Daniel Campos}
\author{Vicen\c{c} M\'endez}
\affiliation{Grup de F{\'i}sica Estad\'{i}stica.  Departament de F{\'i}sica. Facultat de Ci\`{e}ncies. Edifici Cc. Universitat Aut\`{o}noma de Barcelona, 08193 Bellaterra (Barcelona) Spain}
%\ead{axel.maso@uab.cat}

\begin{abstract}
In this work we consider a stochastic movement process with random resets to the origin followed by a random residence time there before the walker restarts its motion. First, we study the transport properties of the walker, we derive an expression for the mean square displacement of the overall process and study its dependence with the statistical properties of the resets, the residence and the movement. From this general formula, we see that the inclusion of the residence after the
resets is able to induce super-diffusive to sub-diffusive (or diffusive) regimes and it can also make a sub-diffusive walker reach a constant
MSD or even collapse. Second, we study how the reset-and-residence mechanism affects the first arrival time of different search processes to
a given position, showing that the long time behavior of the reset and residence time distributions determine the existence of the mean first arrival time.
\end{abstract}
%\keywords{ }
\maketitle

\section{Introduction}

The territorial dynamics of animals are compound multi-stage processes. Their behaviour is completely different when they are seeking for food than when they are resting or socializing around their nest or just exploring potential areas to migrate. Therefore, a complete model to describe how animals behave around their nest should consider a combination of different states and the overall process is the one which ends up determining their region of influence.

Among these states, in the animal ecology context, search processes have been the main object of study ~\cite{OkLe02,BaDa05,MeCaBa14} but the resting times at the nest or migrations \cite{BaKl13} have also been analysed. Nevertheless, all the studies tend to confront each problem independently and the multi-stage perspective is still lacking.

As a model of an animal returning to its nest between multiple excursions, the inclusion of resets to a diffusive motion \cite{EvMa11} is a first step in the formulation of a model cappable of capturing the global dynamics from the multiple internal states of the animals. After that, multiple papers have been devoted to study many types of processes with different resetting mechanisms \cite{EvMa11p, WhEv13, MoVi13, EvMa13, GuMa14, EvMa14, DuHe14, KuMa14, KuGu15, Pa15, MaSa15p, CaMe15, ChSc15, MeCa16, PaKu16, NaGu16, EuMe16, FaEv17, BoEv17, Sh17, Be18}, some of them focusing on the completion time of the processes \cite{RoRe15, Re16, PaRe17, ChSo18}, or using the reset to concatenate different processes \cite{MoVi16,MoMa17}.

However, the vast majority of the existing works treat resets as an instantaneous action that connects two different realisations of a given process, which for most of the real world applications of stochastic search, including movement ecology, is not reallistic. In this work we introduce a residence time during which the walker stays motionless after a reset happens, with the aim of modelling the cooling period before a new excursion begins. For the overall process we compute an analytical expression of its mean square displacement (MSD) and study its first arrival statistics in terms of the type of stochastic movement, the resetting distribution and the residence period. The formalism herein employed resembles the ones we used in \cite{MaCaMe18} to study stochastic processes with resets from a general perspective.

Despite our focus is put on modelling animal foraging dynamics, the process studied in this work can be used to model many other systems. For instance, the residence time after a reset could be seen as the time a stochastic search algorithm needs to be prepared for a new execution or, in a more general perspective, as the time needed to return to the point from which a new search must start.

This paper is organised as follows. In Section II we present the model and we derive an expression for the probability of the walker being at position $x$ at time $t$. In Section III we find a formula for the global MSD in terms of the MSD of the movement process and its long-term behaviour is analysed, while in Section IV we study the case where a stationary state is reached. Finally, in Section V the first arrival statistics of the process are studied and, in the cases where it is finite, a general expression for the mean first arrival time (MFAT) is found. We conclude the work in Section VI.

\section{The Process}
\label{SecProcess}

Let us introduce the general formulation of the process studied in this paper. In order to describe both the movement process and the resting of the walker in the nest, we define two different states. The first state, $i=1$, corresponds to the movement stage of the walker, while in the second state, $i=2$, the walker rests at the origin. Before introducing the spatial behaviour, we study the transition probabilities between the two states. Let $\varphi_R(t)$ be the reset time distribution (i.e. the distribution of the times the walker spends travelling) and $\varphi_S(t)$ the resting or residence time distribution (i.e. the distribution of time that the walker stays at the origin before moving again). Then, the probability of arriving at state $i=1,2$ at time $t$ can be written as

\begin{align}
j_1(t)&=\delta(t)+\int_0^t j_{2}(t-t')\varphi_{S}(t')dt'
\label{Eq1}
\\
j_2(t)&=\int_0^t j_{1}(t-t')\varphi_{R}(t')dt',
\label{Eq2}
\end{align}\\
where the $\delta(t)$ in the first equation indicates that the process starts at state $i=1$. The second term in the first equation and the second equation are the probabilities of reaching the state $i$ from state $i'$; which is the probability of reaching the state $i'$ at any past time $t-t'$ and stay there for a time $t'$, when the walker jumps back to $i$. Transforming Eqs.\eqref{Eq1}\eqref{Eq2} by Laplace for the time variable we get 

\begin{align}
\hat{j}_1(s)&=\frac{1}{1-\hat{\varphi}_R(s)\hat{\varphi}_S(s)}
\label{Eq5}
\\
\hat{j}_2(s)&=\frac{\hat{\varphi}_R(s)}{1-\hat{\varphi}_R(s)\hat{\varphi}_S(s)}.
\label{Eq6}
\end{align}

The next step is to introduce the territorial motion to the model. The overall probability that the walker is at point $x$ at time $t$ can be split into two parts $\rho(x,t)=\rho_1(x,t)+\rho_2(x,t)$, where $\rho_i(x,t)$ is the probability of the walker when it is at state $i$. For the state $i=1$ we define the propagator $P(x,t)$ as the time-dependent distribution of the walker position during a single movement stage. On the other hand, we take the position of the the walker to be fixed at $x=0$ when it is in the resting state ($i=2$).  With these considerations, the time-dependent probability density function (pdf) for each of the states becomes:

\begin{align}
\rho_1(x,t)&=\int_0^t dt' j_1(t-t')\varphi_R^*(t')P(x,t')
\label{Eq7}
\\
\rho_2(x,t)&=\delta(x) \int_0^t dt' j_2(t-t')\varphi_S^*(t')
\label{Eq8}
\end{align}
where $\varphi_{R,S}^*(t)=\int_t^{\infty}\varphi_{R,S}(t')dt'$. These equations can be read as follows: the position distribution of the walker in each of the states $i=1,2$ at time $t$ is the probability of getting there any time before ($j_1(t-t')$ and $j_2(t-t')$ respectively), stay there for the remaining time ($\varphi_R^*(t')$ and $\varphi_S^*(t')$ respectively), during which the dynamics of the walker are described for the corresponding propagator ($P(x,t')$ and $\delta(x)$ respectively). Transforming by Laplace in time, and substituting the transition probabilities by their explicit expressions in Eq.\eqref{Eq5} and Eq.\eqref{Eq6} we get

\begin{align}
\hat{\rho}_1(x,s)&=\frac{\mathcal{L}\left[\varphi_R^*(t)P(x,t)\right]}{1-\hat{\varphi}_R(s)\hat{\varphi}_S(s)}
\label{Eq9}
\\
\hat{\rho}_2(x,s)&= \frac{\hat{\varphi}_S^*(s)\hat{\varphi}_R(s)}{1-\hat{\varphi}_R(s)\hat{\varphi}_S(s)} \delta(x)
\label{Eq10}
\end{align}
where $\mathcal{L}[f(t)]= \hat{f}(s)=\int_0^\infty e^{-st}f(t) dt$. Hence,

\begin{equation}
\hat{\rho}(x,s)= \frac{\hat{\varphi}_S^*(s)\hat{\varphi}_R(s) \delta(x)+\mathcal{L}\left[\varphi_R^*(t)P(x,t)\right]}{1-\hat{\varphi}_R(s)\hat{\varphi}_S(s)}
\label{Eq11}
\end{equation}
is the pdf of the global process. Notably, when the time distribution at the origin is taken as $\varphi_S(t)=\delta(t)$, this expression reduces to the one found in \cite{MaCaMe18}. Also, if we consider only exponentially distributed resets, this expression reduces to the result found in \cite{EvMa18}.

\section{Transport Regime Analysis}

In the previous section we have derived an expression for the probability that the walker is at point $x$ at time $t$. Here we will study the asymptotic behaviour of MSD. Concretely, we study how the tails of the reset and reresidence time distributions modify the transport properties of the global process, leaving for the next section the study of the cases where the walker reaches a stationary state.

To do so, we start from Eq.\eqref{Eq11}. If we multiply at both sides by $x^2$ and integrate over the whole spatial domain we get an equation for the global MSD in the Laplace space in terms of the time distributions and the movement MSD:

\begin{equation}
\mathcal{L}[\langle x^2(t)\rangle ] =\frac{\mathcal{L}\left[ \varphi_R^*(t) \langle x^2(t)\rangle_P \right]}{1-\hat{\varphi}_R(s)\hat{\varphi}_S(s)}.
\label{Eq12}
\end{equation}
Here, the subindex $P$ in $\langle x^2(t)\rangle_P$ indicates that the MSD corresponds to the movement stage (without resets), while the expression without subindex $\langle x^2(t)\rangle$ corresponds to the global process. In order to study more explicitly the MSD we choose a particular expression for the time distributions:

\begin{align}
\varphi_R(t)&=\frac{t^{\gamma_R-1}}{\tau_m^{\gamma_R}}E_{\gamma_R, \gamma_R}\left[ -\left( \frac{t}{\tau_m}\right)^{\gamma_R} \right]
\label{Eq13}
\\
\varphi_S(t)&=\frac{t^{\gamma_S-1}}{\tau_s^{\gamma_S}}E_{\gamma_S, \gamma_S}\left[ -\left( \frac{t}{\tau_s}\right)^{\gamma_S} \right]
\label{Eq14}
\end{align}
with 
\begin{equation}
E_{\alpha, \beta}(z)=\sum_{n=0}^{\infty}\frac{(-z)^n}{\Gamma(\alpha n + \beta)},
\label{Eq15}
\end{equation}
being the Mittag-Leffler function, and $0<\gamma_i\leq 1$ and $\tau_i$ for $i=R,S$ are characteristic parameters. In the large time limit and for $\gamma_i <1$ we have that $\varphi_i(t)\sim t^{-1-\gamma_i}$. The choice of this distribution is motivated by the fact that we can recover exponential reset times for $\gamma_R=1$ and we can also study long tail behaviours for $\gamma_R<1$. If the movement process has a MSD of the form

\begin{equation}
\langle x^2(t)\rangle_P \sim t^p,
\label{Eq16}
\end{equation}
with $0<p<2$, the asymptotic behaviour of the global process MSD can be found by taking Eq.\eqref{Eq12} to small $s$ and performing the inverse Laplace transform on the result. Doing so for the whole range of asymptotic parameters, we have three different behaviours depending on the relative values of $\gamma_R$ and $\gamma_S$:

\begin{equation}
\langle x^2(t)\rangle \sim t^a
\label{Eq17_0}
\end{equation}
with
\begin{equation}
a= \left\{ 
\begin{matrix}
\gamma_S -1&,\gamma_S\leq \gamma_R=1\\
p+\gamma_S -\gamma_R&,\gamma_S<\gamma_R<1\\
p &,\gamma_S \geq \gamma_R<1\\
\end{matrix}
\right.
\label{Eq17}
\end{equation}

The inclusion of the residence at the origin generates a competition between the tails of the distributions $\varphi_R(t)$ and $\varphi_S(t)$, which produces a rich diversity of transport regimes. In the following we study the different cases of Eq.\eqref{Eq17} where the transport regime is modified by the reset-and-residence mechanism, which include long tail and exponential reset time distributions.

\subsection{Long tail $\varphi_R(t)$ distribution}

\begin{figure}
\includegraphics[scale=0.63]{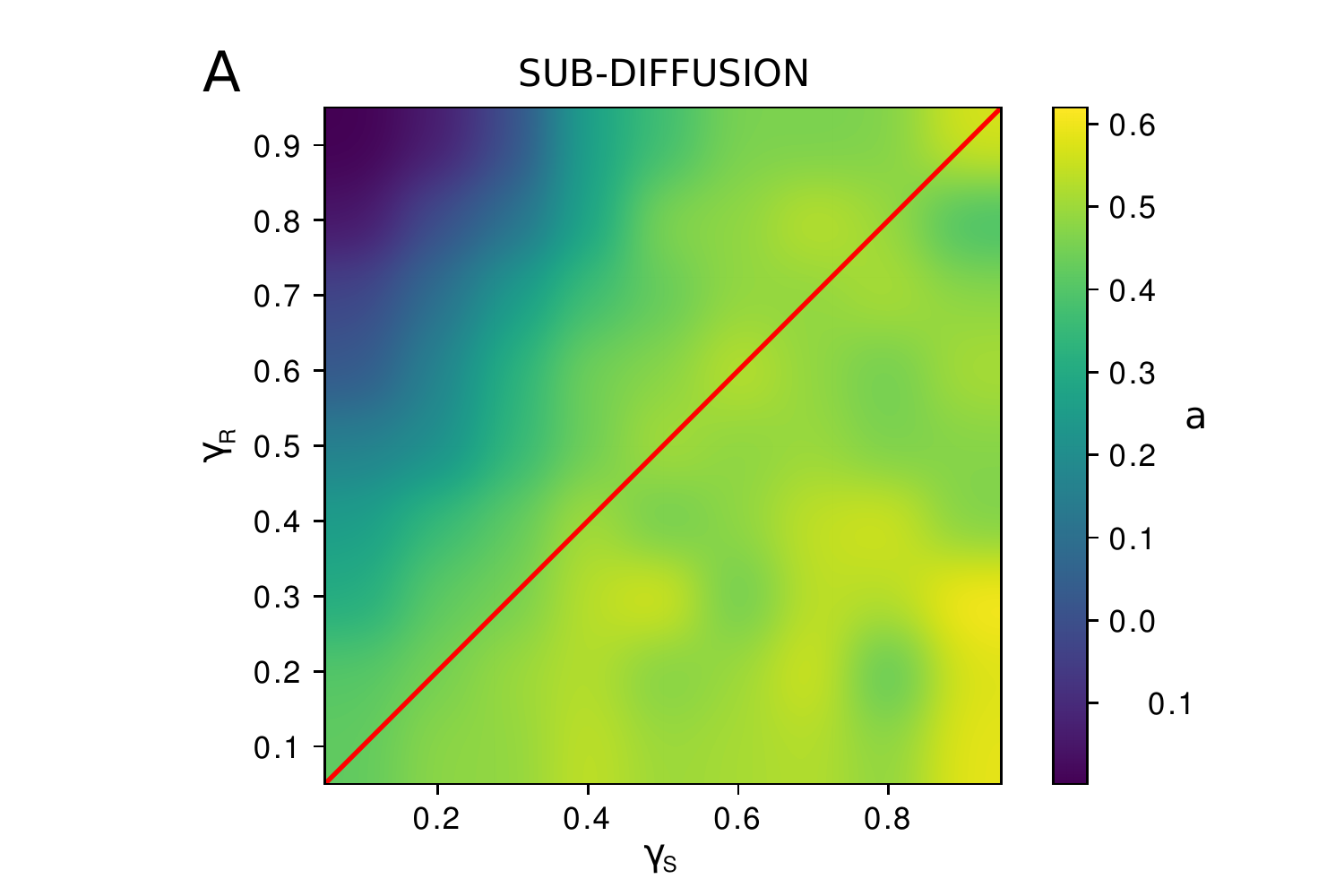}
\includegraphics[scale=0.63]{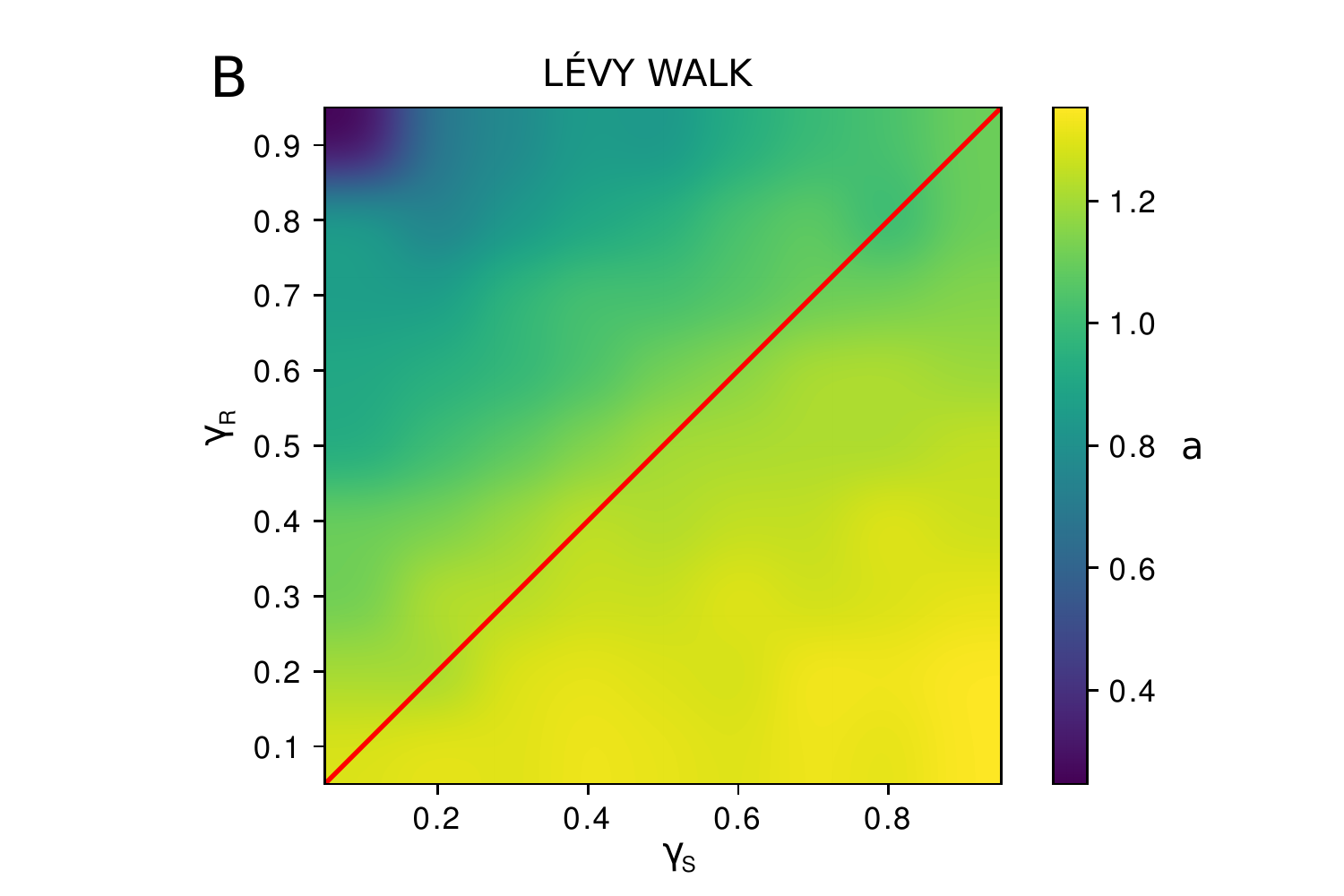}
\caption{Exponent of the asymptotic behaviour of the MSD of the overall process $\langle x^2(t)\rangle\sim t^a$ with long tailed reset and residence time distributions. Their exponents are $\gamma_R$ and $\gamma_S$ and the plots are for simulations of two different movement processes: a sub-diffusive process with $p=0.5$ in A and a super-diffusive L\'evy walk with $p=1.5$ in B. For both cases, in the region below the red line ($\gamma_R=\gamma_S$) the exponent is constant and equal to the movement MSD exponent $a=p$ while in the region above the line the exponent diminishes progressively.}
\label{Fig1}
\end{figure}

Let us start with the two last cases, where the reset time distribution has a long tail. In Fig.\ref{Fig1} the exponent of the MSD at the asymptotic limit, obtained with Monte Carlo simulation, is shown for the whole range of $0<\gamma_S<1$ and $0<\gamma_R<1$, for both a sub-diffusive and a super-diffusive case. There we can observe the two bottom cases in Eq.\eqref{Eq17}, separated by the analytical limiting case $\gamma_R=\gamma_S$, represented by a red line. Below the red line the asymptotic exponent of the MSD does not depend on the values of $\gamma_R$ and $\gamma_S$ as in the third case while above the red line the asymptotic exponent decreases by a factor $\gamma_R-\gamma_S$ as we have actually derived in the second case of Eq.\eqref{Eq17}. 

Interestingly, the conditions for the different cases in Eq.\eqref{Eq17} do not depend on the movement transport regime exponent $p$. Therefore, the competition is exclusively between the tails of the time distributions, $\gamma_R$ and $\gamma_S$. When the tail of $\varphi_R(t)$ decays equal or slower than the tail of $\varphi_S(t)$ ($\gamma_S \geq \gamma_R<1$), at long times there are less walkers resetting their position than leaving the origin, which makes the residence stage negligible. On the other hand, when the tail of $\varphi_S(t)$ decays slower than the tail of $\varphi_R(t)$ ($\gamma_S<\gamma_R<1$), the residence distribution becomes significant and the overall transport regime is affected by the reset and residence time distributions. In fact, this change of behaviour is a consequence of the requirement of the walker to have restarted its position for the residence mechanism to be triggered. Therefore, when the reset times are asymptotically longer than the residence times, the asymptotic behaviour of the MSD is as if there was no residence \cite{MaCaMe18}. Otherwise, when the residence times are asymptotically longer, the effect of the particles stacked at the origin becomes qualitativaly relevant. In Fig.\ref{Fig1} we show the results obtained from Monte-Carlo simulations of different movement processes, which are in agreement with our analytical prediction.

\subsection{Exponential $\varphi_R(t)$ distribution}

\begin{figure}
\includegraphics[scale=0.6]{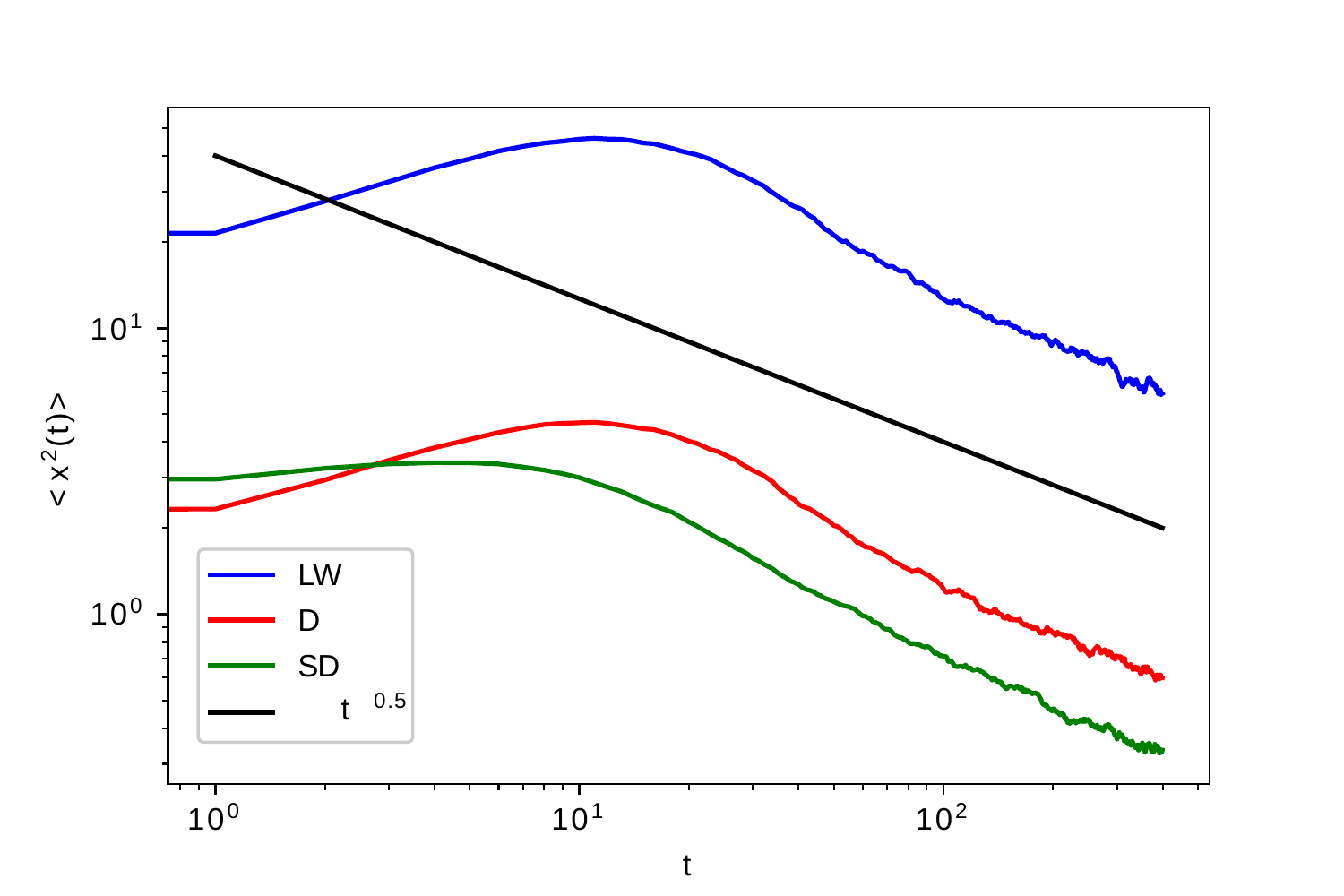}
\caption{The time evolution of the MSD of a simulation of an overall process with $\gamma_R=1$ and $\gamma_S=0.5$ is plotted for a sub-diffusive (SD), a diffusive (D) and a L\'evy walk (LW) super-diffusive motion. In all three cases the MSD decreases as $t^{-0.5}$ as predicted by the first case in Eq.\eqref{Eq17} as shown with the guide line (black line in the plot).}
\label{Fig2}
\end{figure}

When the reset distribution $\varphi_R(t)$ is exponential (first case in Eq.\eqref{Eq17}) the asymptotic behaviour of the MSD changes drastically. This case includes both power law and exponential distributions for $\varphi_S(t)$. In Fig.\ref{Fig2} we show the simulated MSD for long tailed residence time distributions and different movement: sub-diffusive and diffusive processes, and super-diffusive L\'evy walks. In all three cases the MSD increases for short times until it reaches a maximum value and starts to collapse. The asymptotic decaying exponent depends exclusively on the residence time distribution exponent $\gamma_S$ and in this regime the movement process only modifies multiplicatively the global MSD.

$ $\\

\begin{figure}
\includegraphics[scale=0.6]{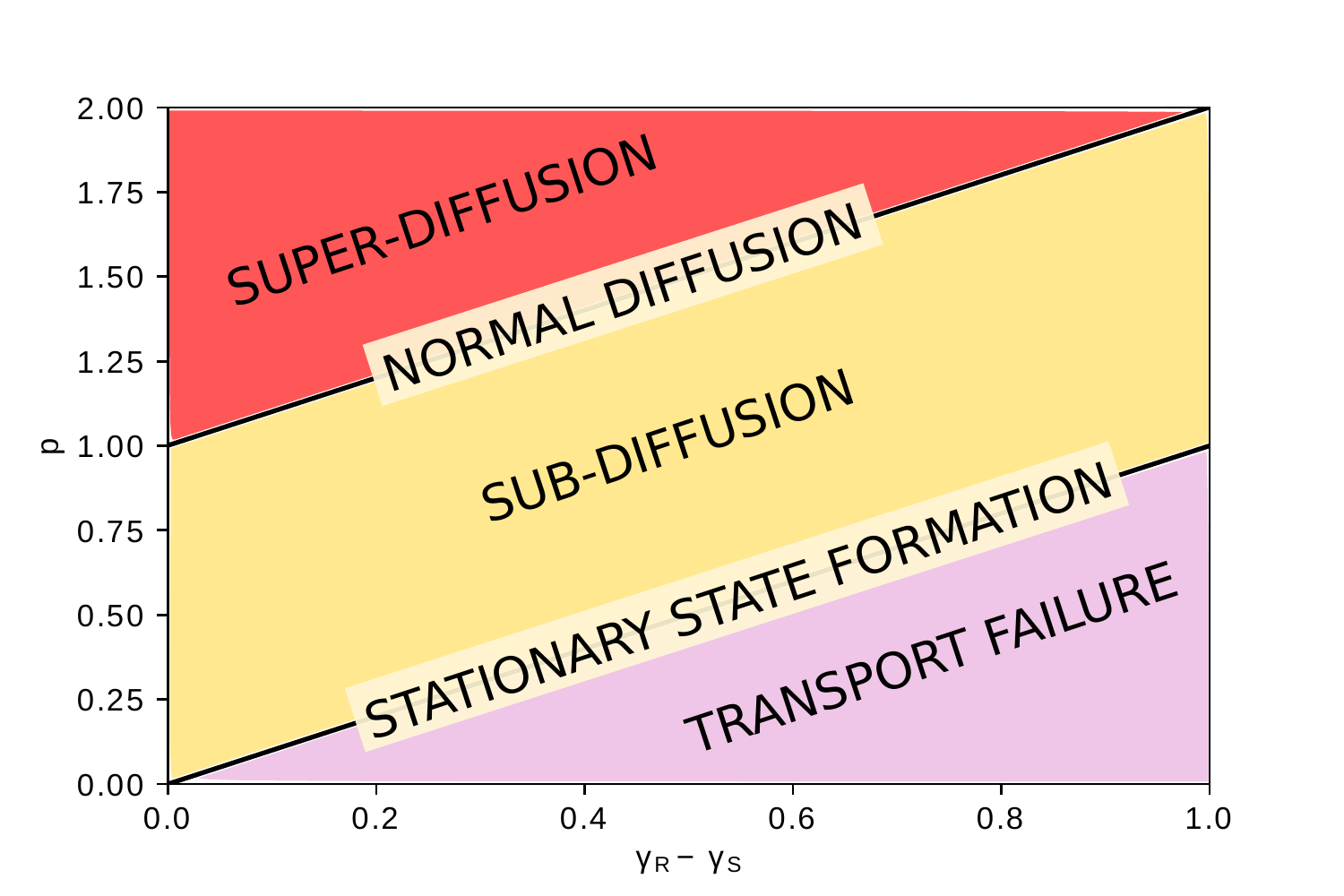}
\caption{Schematic plot of the different global transport regimes in terms of the transport regime of the movement process $p$ and the difference of the decaying exponents of the reset and retention time distributions $\gamma_R-\gamma_S$. In the red region the overall behaviour is super-diffusive. in the yellow region it is sub-diffusive and the pink represents the values for which we have transport failure.}
\label{Fig1_2}
\end{figure}

In Fig.\ref{Fig1_2} we present a summary of the possible transport regimes. There we can see that a super-diffusive process ($p>1$) can be transformed to a sub-diffusive (or diffusive) process when the reset-and-residence mechanism is applied to it. Similarly, a sub-diffusive process ($p<1$) can derive into a transport failure regime when the joint mechanism is applied. Therefore, the reset-and-residence mechanism can be also seen as a tool to tune the transport regime of a certain process.  

\section{Stationary State}

In the previous section we have derived in a general manner the transport regime of the global process and we have also studied how the reset-and-retention mechanism affects the long term behaviour of the system. In this section we study the special case where both the residence and the reset time distributions are exponential ($\gamma_S=\gamma_R=1$), when a stationary state is reached. Concretely, we derive a formula for the global MSD in terms of the MSD of the movement and we find the stationary distribution for some well-known movement processes. 

To do so, we start from Eq.\eqref{Eq12} and the time distributions from Eq.\eqref{Eq13} and Eq.\eqref{Eq14} with $\gamma_R=1$ and $\gamma_S=1$ respectively. With these parameters the Laplace transform of the global MSD reads from Eq.\eqref{Eq12}

\begin{equation}
\mathcal{L}[\langle x^2(t)\rangle]=\frac{\mathcal{L}[e^{-\frac{t}{\tau_m}}\langle x^2(t)\rangle_P]}{1-\frac{1}{(1+\tau_m s)(1+\tau_s s)}}.
\label{Eq18}
\end{equation}

In order to find the asymptotic behaviour of the MSD we make use of the final value theorem $\lim_{t\rightarrow \infty}f(t)=\lim_{s\rightarrow 0} s \mathcal{L}[f(t)]$. Doing so, the stationary MSD can be easily found from Eq.\eqref{Eq18} giving

\begin{equation}
\langle x^2\rangle_{st}=\frac{1}{\tau_m+\tau_s}\int_0^{\infty}e^{-\frac{t}{\tau_m}}\langle x^2(t) \rangle_{P}\ dt
\label{Eq19}
\end{equation}
which is valid for any motion whose MSD as a function of time is Laplace-transformable. For instance, for a process with $\langle x^2(t)\rangle_P= 2 D t^p$ with $D,\ p$ positive constants, the stationary MSD takes the simple form:

\begin{equation}
\langle x^2\rangle_{st}=2D\ \Gamma(p+1)\frac{\tau_m^{p+1}}{\tau_m+\tau_s}.
\label{Eq20}
\end{equation}

\begin{figure}
\includegraphics[scale=0.6]{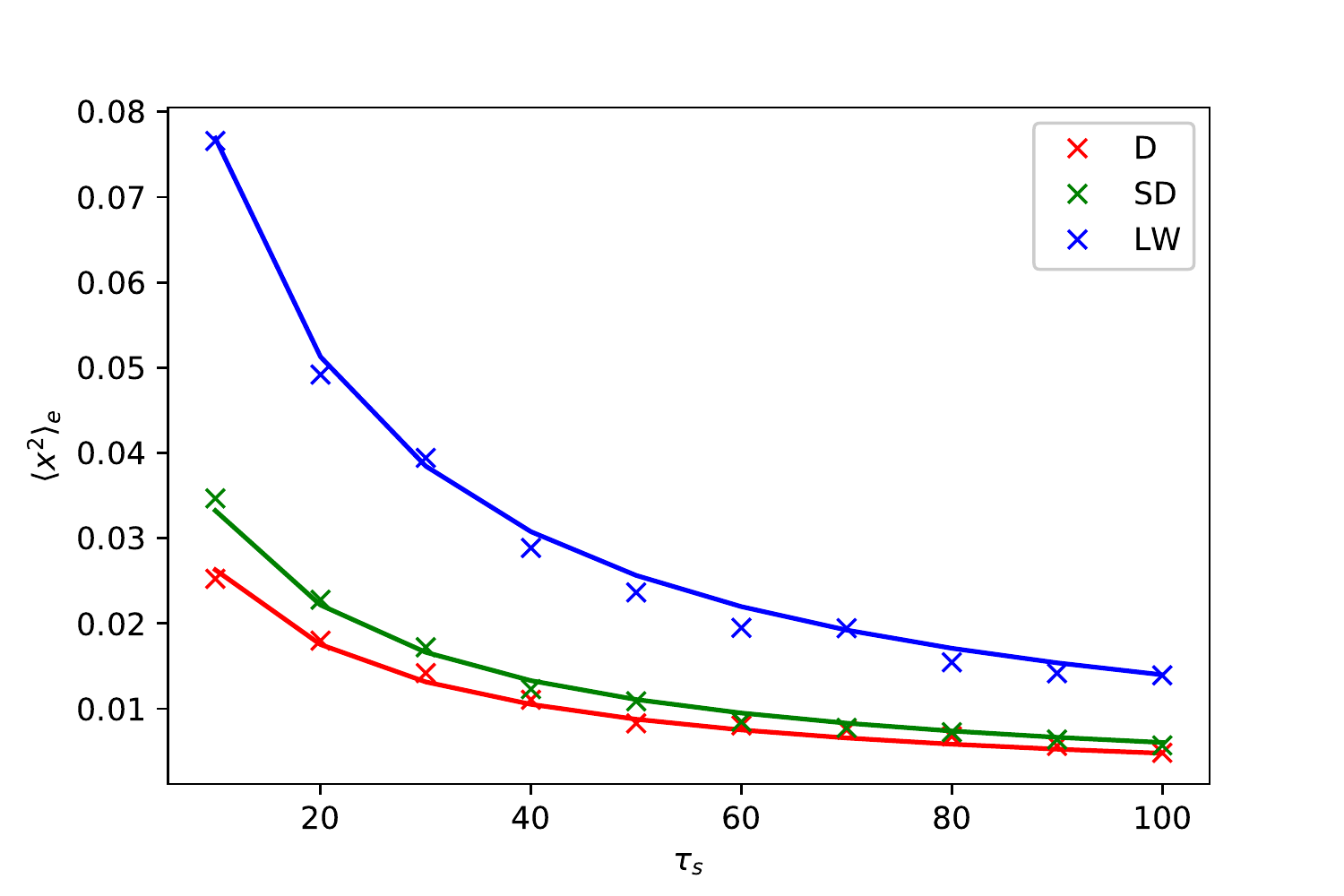}
\caption{Stationary MSD for a sub-diffusive (SD), a diffusive (D) and a L\'evy flight (LF) propagator when they are subject to reset times which are exponential distributed with $\tau_m=10$ and retained at the origin during a time given by an exponential distribution with different means $\tau_s$. For the propagators we have chosen $p=0.5,\ D=1.05\cdot 10^{-2}$ for the sub-diffusive, $p=1,\ D=2.63\cdot 10^{-3} $ for the diffusive and $p=1.1,\ D= 6.11\cdot 10^{-3}$ for the L\'evy walk. The crosses correspond to the simulated distributions while the solid curves show the analytical result in Eq.\eqref{Eq20}.}
\label{Fig3}
\end{figure}

This result has been compared with Monte Carlo simulations of the process in Fig.\ref{Fig3} for three different types of movement. Let us calculate the complete stationary distribution for some cases of interest by proceeding as with the MSD but starting from Eq.\eqref{Eq11} instead of Eq.\eqref{Eq12}. Doing so it is easy to see that the global stationary distribution can be written in terms of the propagator as

\begin{equation}
\rho_{st}(x)=\frac{\tau_s}{\tau_m+\tau_s}\delta(x)+\frac{\hat{P}(x,\frac{1}{\tau_m})}{\tau_m+\tau_s}.
\label{Eq21}
\end{equation}

For a propagator of the form $P^{SD}(k,s)=\frac{1}{s+D_{\gamma} s^{1-\gamma}k^2}$ in the Fourier-Laplace space, which corresponds to a sub-diffusive process for $\gamma<1$ and diffusive for $\gamma=1$, the global stationary distribution is

\begin{equation}
\rho_{st}^{SD}(x)=\frac{\tau_s}{\tau_m+\tau_s}\delta(x)+\frac{\tau_m}{\tau_m+\tau_s} \frac{e^{-\frac{|x|}{\sqrt{D_{\gamma} \tau_m}}}}{\sqrt{4D_{\gamma}\tau_m^{\gamma}}}.
\label{Eq22}
\end{equation}

If instead of a sub-diffusive (or diffusive) motion we have a L\'evy Flight with propagator $P^{LF}(k,s)=\frac{1}{s+D_{\alpha} |k|^{\alpha}}$ in the Fourier-Laplace space, where $\alpha<2$ and $D_{\alpha}$ is the corresponding diffusion constant, the stationary distribution reads 

\begin{equation}
\rho_{st}^{LF}(x)=\frac{\tau_s}{\tau_m+\tau_s}\delta(x)+\frac{2 \tau_m}{\tau_m+\tau_s} \int_0^{\infty} \frac{\cos(kx)}{1+\tau_m D_{\alpha} k^{\alpha}}dk.
\label{Eq23}
\end{equation}

This distribution has infinite second order moment so it is a power law for large $|x|$. Therefore, despite a stationary state is reached, the average region of space covered by the walker around the nest is infinite. Contrarily, for the sub-diffusive and diffusive movement processes we have a finite stationary MSD, the value of which is given by Eq.\eqref{Eq20} with properly chosen parameters. 

\section{Mean First Arrival Time}

In the previous section we have analysed the long term behaviour of the overall process. Instead of this, in the following we study the first arrival time properties of the model. More concretely, we determine the conditions for which the MFAT is finite. To do so, we build a mesoscopic equation for the global survival probability $\sigma_x(t)$, which is the probability of the walker not having arrived to $x$ at time $t$. Let us first introduce three mutually excluent cases which we will use later to write down the matematical equation for $\sigma_x(t)$. If the walker starts at the origin but is moving (i.e. in state $i=1$), the three contributions to the global survival probability are:

\begin{itemize}
\item[i)] The walker has not yet restarted at time $t$. This happens with probability $\varphi_R^*(t)$ and in this case the global survival probability is the same as the motion survival probability $Q_x(t)$. This scenario contributes as the first term in Eq.\eqref{Eq24}.

\item[ii)] The walker has restarted its position once at time $t'<t$, during which it has not arrived to $x$, and during the period $(t',t)$ it has not left the origin. The first happens with probability $\varphi_R(t')Q_x(t')dt'$, the second with probability $\varphi_S^*(t-t')$ and $t'$ can be any time from $0$ to $t$. Here the mathematical expression is the one in the second term from Eq.\eqref{Eq24}.

\item[iii)] The walker has restarted its position at least once at time $t'<t$ and it has not reached $x$ before $t'$, which happens with probability $\varphi_R(t')Q_x(t')dt'$ as in the previous case. However, here the walker leaves the origin after a time $t''<t-t'$ and in this case the walker is exactly at the initial scenario but instead of having a time $t$, we have to subtract the time of the first non-triggering journey ($t\rightarrow t-t'-t''$). This happens with probability $\varphi_S(t'')\sigma_x(t-t'-t'')dt''$ given that the first reset was at time $t'$. These scenarios must be taken into account for all $t''<t-t'$ and $t'<t$. This situation corresponds to the last term in Eq.\eqref{Eq24}.
\end{itemize}

Putting the three cases in mathematical form we get:

\begin{equation}
\begin{split}
\sigma_x(t)= &\ \varphi_R^*(t)Q_x(t)+\int_0^t \varphi_R(t')Q_x(t')\varphi_S^* (t-t')dt'\\
+ & \int_0^t \varphi_R(t')Q_x(t')dt'\int_0^{t-t'}\varphi_S(t'')\sigma_x(t-t'-t'')dt'',
\end{split}
\label{Eq24}
\end{equation}
where $Q_x(t)$ is the survival probability of the movement process. Applying the Laplace transform to Eq.\eqref{Eq24} we get

\begin{equation}
\hat{\sigma}_x(s)=\frac{\mathcal{L}[\varphi_R^*(t)Q_x(t)]+\mathcal{L}[\varphi_R(t)Q_x(t)]\hat{\varphi}_S^*(s)}{1-\mathcal{L}[\varphi_R(t)Q_x(t)]\hat{\varphi}_S(s)}. 
\label{Eq25}
\end{equation}

This equation is completely general so it is valid for any survival probability of the movement process $Q_x(t)$ and any reset and residence time distributions $\varphi_R(t)$ and $\varphi_S(t)$. In order to get an intuition about the asymptotic behaviour of $\sigma_x(t)$, which determines the existence of the MFAT, we make some assumptions on the asymptotic form of the fundamental distributions of the model. Regarding the motion survival probability we restrict to the vast class of cases where $Q_x(t) \sim t^{-q}$ with $0<q<1$ as $t\rightarrow \infty$. For $\varphi_R(t)$ and $\varphi_S(t)$ we consider Eq.\eqref{Eq13} and Eq.\eqref{Eq14} respectively. 

Now, depending on the tail exponents of the distributions $\varphi_R(t)$ and $\varphi_S(t)$, the dominant term in the $t\rightarrow \infty$ limit of Eq.\eqref{Eq25} can be the first or the second term in the numerator. The existence of a MFAT can be determined by taking the limit $s\rightarrow 0$ in Eq.\eqref{Eq25} ($T_F=\lim_{s\rightarrow 0}\hat{\sigma}(s)$). Here we present the cases that appear for different tails:

\begin{itemize}
\item[a)] $\gamma_S \geq \gamma_R + q$. At long times, the first term in the numerator of Eq.\eqref{Eq25} dominates over the second (or equal in the limiting case). In this case, the MFAT is infinite since the survival probability decays as

\begin{equation}
\sigma_x(t)\sim t^{\gamma_R+q-1}
\label{Eq26}
\end{equation}

\item[b)] $\gamma_S<\gamma_R+q$, $\gamma_S<1$. At long times, the second term in the numerator of Eq.\eqref{Eq25} is the dominant term. Here, the MFAT is infinite again, with a survival probability decaying as

\begin{equation}
\sigma_x(t)\sim t^{\gamma_S-1}
\label{Eq27}
\end{equation}

\item[c)] $\gamma_S<\gamma_R+q$, $\gamma_S=1$. In this scenario, the MFAT exists. It can be written as

\begin{equation}
T_{F}=\frac{I_1+\tau_s I_2}{1-I_2}
\label{Eq28}
\end{equation}
with
\begin{equation}
I_1=\int_0^{\infty} E_{\gamma_R, 1}\left[-\left( \frac{t}{\tau_m}\right)^{\gamma_R}\right]Q_x(t)dt
\label{Eq29}
\end{equation}
and
\begin{equation}
I_2=\int_0^{\infty} \frac{t^{\gamma_R-1}}{\tau_m^{\gamma_R}}E_{\gamma_R, \gamma_R}\left[ -\left( \frac{t}{\tau_m}\right)^{\gamma_R}\right]Q_x(t)dt,
\label{Eq30}
\end{equation}
where we have made use of Eq.\eqref{Eq13} and Eq.\eqref{Eq14}.
\\
\end{itemize}

Then, when studying the overall survival probability, the reset time distribution is coupled to the survival probability of the movement process when competing with the residence time distribution. Therefore, the type of movement has a direct effect on which distribution determines the tail exponent of the overall survival probability. To ilustrate this, we study deeply these general results for some particular cases.

\subsection{Infinite MFAT}

\begin{figure}
\includegraphics[scale=0.63]{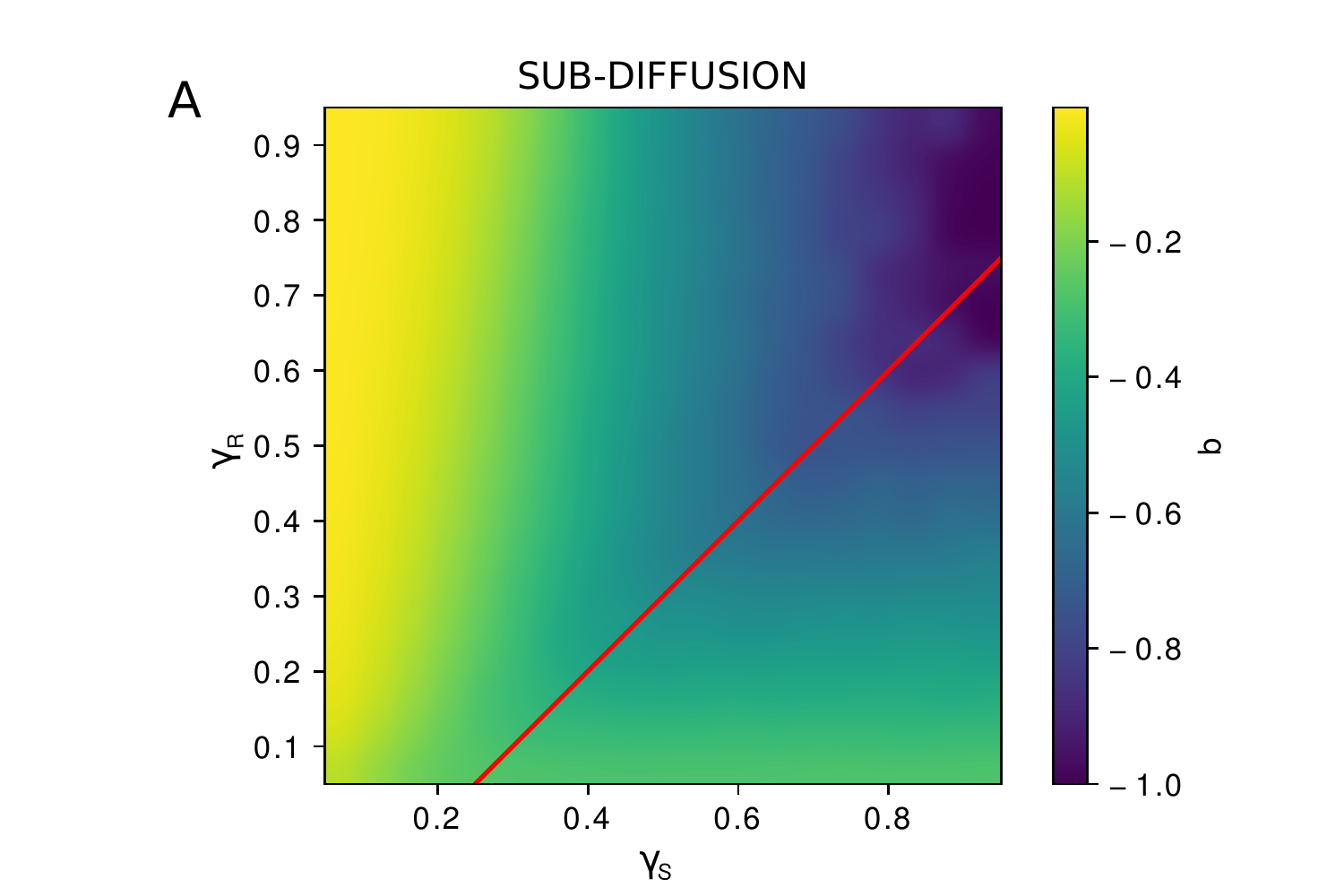}
\includegraphics[scale=0.63]{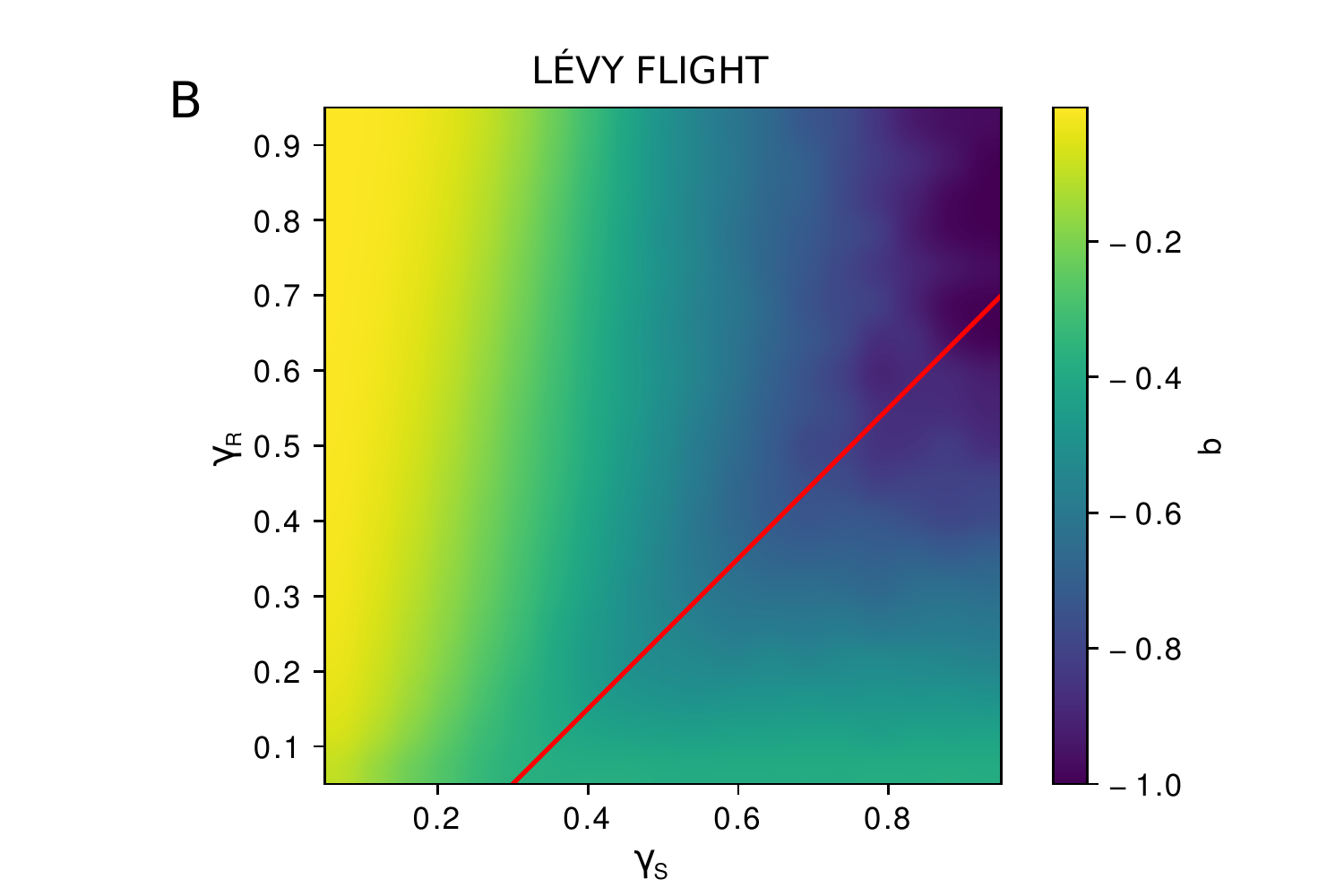}
\caption{Exponent of the asymptotic behaviour of the survival probability of the overall process $\sigma_x(t)\sim t^{-b}$ with long tailed reset and residence time distributions. Their exponents are $\gamma_R$ and $\gamma_S$ and the plots are for simulations of two different movement processes: a sub-diffusive process with $\gamma=0.5$ in A and a super-diffusive L\'evy walk with $\alpha=1.5$ in B. For both cases, in the region above the red line ($\gamma_S=\gamma_R+q$ with $q=0.25$ in A and $q=0.33$ in B) $b$ only changes horizontally with $\gamma_S$ while below the line $b$ only varies vertically with $\gamma_R$. These are the results predicted by the cases a) and b) in the main text respectivelly.}
\label{Fig4}
\end{figure}

Let us leave the case c) apart for the moment. From cases a) and b) we see that there are two different regions in the state space where the governing tail differs, with limit in the plane $\gamma_S=\gamma_R+q$. It must be emphasized that the reset distribution couples its tail with the survival probability of the motion and the coupled exponent $\gamma_R+q$ is the one that competes with the residence time distribution tail exponent $\gamma_S$. Here we test this property for two different movement processes: a sub-diffusive random walk (Fig.\ref{Fig4}A) and a L\'evy flight (Fig.\ref{Fig4}B). 

The survival probability of a sub-diffusive jump process has been studied in \cite{RaDi00}, where it is found to be

\begin{equation}
Q_x^{SD}(t)=\frac{x}{\sqrt{Dt^{\gamma}}} \sum_{k=0}^{\infty} \frac{\left( -\frac{x}{\sqrt{Dt^{\gamma}}}\right)^k}{(k+1)!\Gamma\left( 1-\frac{\gamma}{2}-k\frac{\gamma}{2}\right)}
\label{Eq31_0}
\end{equation}

with $0<\gamma<1$. This, in the large time limit decays as
\begin{equation}
Q_x^{SD}\sim t^{-\frac{\gamma}{2}}.
\label{Eq31}
\end{equation}

For $\gamma=1$ the result recovers the well-known decay for the survival probability of a diffusive process. From the asymptotic expression we can identify $q=\frac{\gamma}{2}$ and therefore in this case the limiting plane is $\gamma_S=\gamma_R+\frac{\gamma}{2}$, which is illustrated as a red line in Fig.\ref{Fig4}A in the particular case $\gamma=0.5$. There we can see that below the line the tail exponent of the global survival probability increases upwards and remains constant horizontally, which means that it depends on the reset tail only. Contrarily, above the red line the decaying exponent of the global survival probability increases from left to right, remaining constant vertically and therefore in this case it only depends on the residence tail; in agreement with cases a) and b) above. 

Regarding the L\'evy flight process, an expression for the Laplace transform of the first arrival survival probability was derived in \cite{ChMe03}:

\begin{equation}
\hat{Q}_x^{LF}(s)=\frac{\alpha D^{1/\alpha}\sin(\pi/\alpha)}{\pi s^{1/\alpha}} \int_0^{\infty} \frac{1-\cos(k x)}{s+D k^{\alpha}} dk
\label{Eq32_0}
\end{equation}
which behaves as

\begin{equation}
Q_x^{LF}(t)\sim t^{\frac{1}{\alpha}-1}
\label{Eq32}
\end{equation}
as $t\rightarrow \infty$. Here $1<\alpha <2$, being $\alpha=2$ the diffusive limit. In this case the limiting surface is $\gamma_S=\gamma_R+\frac{1}{\alpha}$, which has been plotted for the particular value $\alpha=1.5$ in Fig.\ref{Fig4}B as a red straight line. As for the previous example, the line separates two different regions in which the dominant exponent is $\gamma_S$ above the limit and $\gamma_R$ below. 

\subsection{Finite MFAT}

\begin{figure}
\includegraphics[scale=0.6]{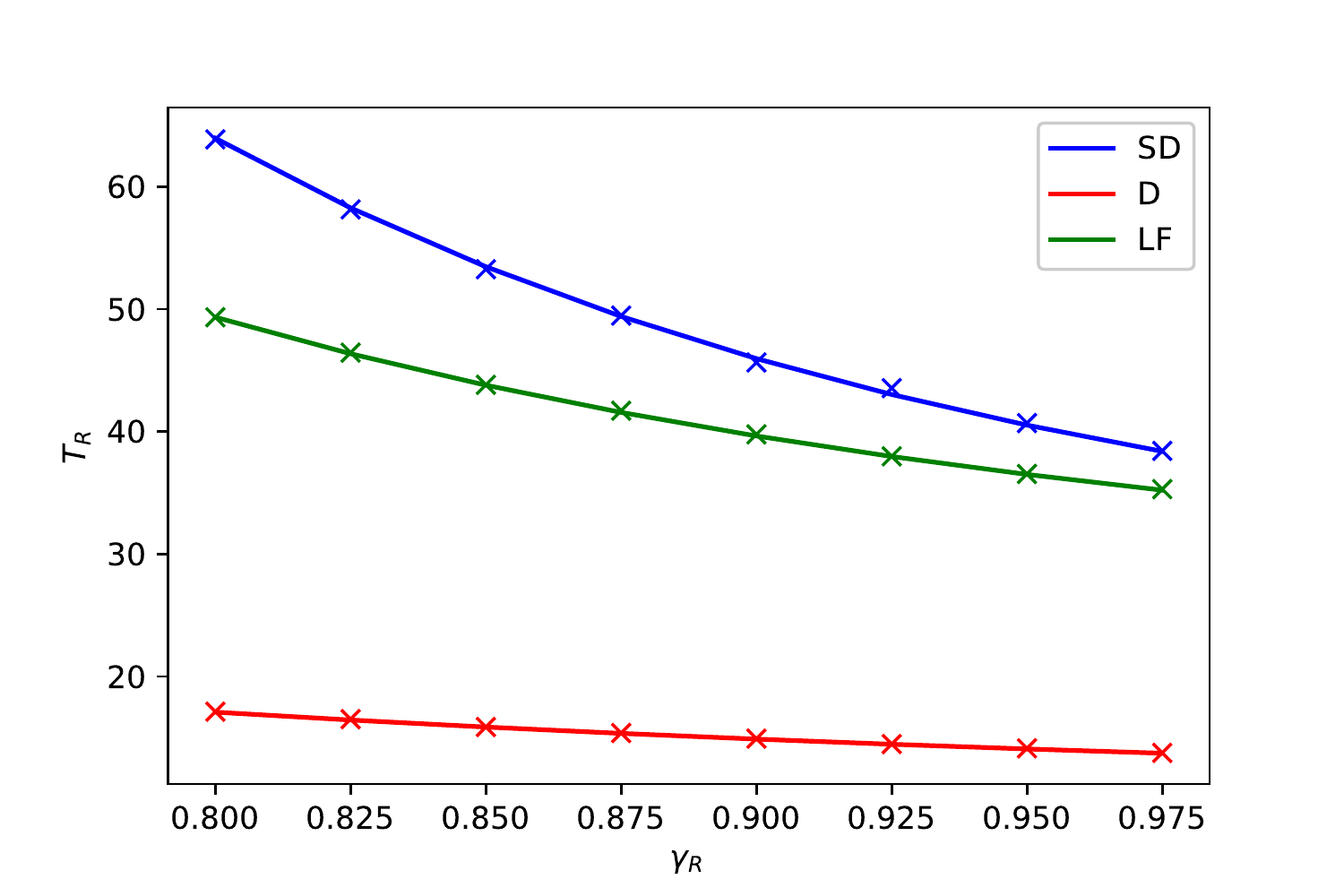}
\caption{The simulated MFAT (crosses) with an exponential residence time distribution with mean $\tau_s=10$ and for three representative cases of subdiffusion (SD) with $\gamma=0.5$, diffusion (D) and a L\'evy Flight with $\alpha=1.5$, all with diffusion constant $D=0.1$, is compared to the analytical result in Eq.\eqref{Eq15} (solid curves) for different reset distribution tail exponents $\gamma_R$ and $\tau_m=10$ for all of them. Concretely, the MFAT is computed at a distance $x=0.5$ from the origin.}
\label{FigMFAT}
\end{figure}

From the three different cases of Eq.\eqref{Eq25}, the c) scenario differs from the others in the fact that the MFAT is finite. In this case, Eq.\eqref{Eq28} gives us its value for different movement survival probabilities. As for the previous cases, we consider the two cases where the motion is a sub-diffusive jump process and a L\'evy flight, the survival probabilities of which are in Eq.\eqref{Eq31_0} and Eq.\eqref{Eq32_0} respectively. In Fig.\ref{FigMFAT} we show the simulated MFAT for sub-diffusive, diffusive and super-diffusive (L\'evy flights) processes in scenarios where it is finite and we compare them with the analytical result obtained by numerically integrate Eq.\eqref{Eq28} for each one of the different foraging processes. 

\section{Conclusions}

We have studied how a stochastic residence period at the resetting position affects the overall dynamics of the walker. More specifically, we have employed a renewal approach to derive analytical expressions for the MSD and the survival probability (or MFAT when it is finite), which have been validated with Monte Carlo simulations of different processes as diffusion, L\'evy Walks or L\'evy Flights. 

For the MSD we have found that, on one hand, for long tailed reset and residence time distributions the diffusivity of the overall process depends strongly on the relation between the tail exponents of both distributions ($\gamma_R$ for resets and $\gamma_S$ for residence times). Concretely, for $\gamma_R\geq \gamma_S$ the overall MSD exponent is the same as the one for the movement process while for $\gamma_R>\gamma_S$ the exponent of the movement process is diminished by a factor $\gamma_R-\gamma_S$. On the other hand, for exponentially distributed reset times, if the residence time distribution is long tailed the overall process collapses to the origin while for an exponential residence time distribution a stationary MSD is reached (see Eq.\eqref{Eq19}).

Regarding the decaying exponent of the survival probability when both time distributions are long tailed, we have seen that the tail of the residence time distribution competes with both the tail of the movement process survival probability and the tail of the reset time distribution. In this case the MFAT is always infinite. Contrarily, when the residence time distribution is exponential, a finite MFAT may be found when $q+\gamma_R >1$ with $q$ and $\gamma_R$ being the tail exponents of the survival probability of the movement process and the reset time distribution respectively.

In the context of movement ecology, resets are a natural mechanism to concatenate different trials of a particular behaviour of the animals. In this work we have included the period that they spend in the nest separating excursions for feeding, mating or any other basic function. A next step on the description of real trajectories could be the introduction of multiple search strategies, generating different movement patterns, to be chosen during the resting period, similarly to what has been done in \cite{MoVi16, MoMa17} for instance.

\bibliography{Referencesa}

\end{document}